\documentclass[iop,revtex4]{emulateapj}
\usepackage{graphicx}
\usepackage[colorlinks,urlcolor=red,citecolor=blue,linkcolor=cyan]{hyperref} 
\pdfoutput=1

\begin{document}
\slugcomment{ApJL, 810, L23}

\shorttitle{Nodal Precession of WASP-33 \lowercase{b}}
\shortauthors{Johnson et al.}

\title{Measurement of the Nodal Precession of WASP-33 \lowercase{b} via Doppler Tomography}

\author{Marshall C. Johnson\altaffilmark{1}, William D. Cochran\altaffilmark{1}, Andrew Collier Cameron\altaffilmark{2}, and Daniel Bayliss\altaffilmark{3} }

\altaffiltext{1}{Department of Astronomy and McDonald Observatory, University of Texas at Austin, 2515 Speedway, Stop C1400, Austin, TX 78712, USA; mjohnson@astro.as.utexas.edu}
\altaffiltext{2}{SUPA, School of Physics \& Astronomy, University of St.~Andrews, North Haugh, St.~Andrews, Fife, KY16 9SS, UK}
\altaffiltext{3}{Observatoire Astronomique de l'Universit\`e de Gen\'eve, 51 ch. des Maillettes, 1290 Versoix, Switzerland}

\begin{abstract}

We have analyzed new and archival time series spectra taken six years apart during transits of the hot Jupiter WASP-33 b, and spectroscopically resolved the line profile perturbation caused by the Rossiter-McLaughlin effect. The motion of this line profile perturbation is determined by the path of the planet across the stellar disk, which we show to have changed between the two epochs due to nodal precession of the planetary orbit. We measured rates of change of the impact parameter and the sky-projected spin-orbit misalignment of $db/dt=-0.0228_{-0.0018}^{+0.0050}$ yr$^{-1}$ and $d\lambda/dt=-0.487_{-0.076}^{+0.089}$~$^{\circ}$ yr$^{-1}$, respectively, corresponding to a rate of nodal precession of $d\Omega/dt=0.373_{-0.083}^{+0.031}$~$^{\circ}$ yr$^{-1}$. This is only the second measurement of nodal precession for a confirmed exoplanet transiting a single star. Finally, we used the rate of precession to set limits on the stellar gravitational quadrupole moment of $9.4\times10^{-5}<J_2<6.1\times10^{-4}$. 

\end{abstract}


\section{Introduction}

WASP-33 b is a hot Jupiter orbiting a relatively massive ($1.5 M_{\odot}$), rapidly rotating ($v\sin i_{\star}=85.6$ km s$^{-1}$) star \citep{CollierCameron10}, which is notable for being one of the hottest planet host stars known ($T_{\mathrm{eff}}=7430$ K). Due to the wide, rotationally broadened stellar lines, \cite{CollierCameron10} were only able to set an upper limit on the radial velocity reflex motion of the host star due to the planet. Even in the \emph{Kepler} era, detection of this motion is typically necessary to confirm a transiting giant planet candidate as a {\it bona fide} planet. Instead, they confirmed the planetary nature of the transiting companion using Doppler tomography. This method relies upon the Rossiter-McLaughlin effect \citep{Rossiter24,McLaughlin24}, where the transit of a companion across a rotating star causes a perturbation to the rotationally broadened stellar line profile. Doppler tomographic observations allow us to resolve this line profile perturbation spectroscopically, unlike typical radial velocity Rossiter-McLaughlin observations, where the perturbation is interpreted as an anomalous radial velocity shift due to the changing photocenter of the line \citep[e.g.,][]{Triaud10}. 
 Most importantly for this work, the movement of the line profile perturbation across the line profile during the transit maps directly to the path of the planet across the stellar disk, allowing us to measure the location and orientation of the transit chord relative to the projected stellar rotation axis. 

Using their Doppler tomographic observations of WASP-33 b, \cite{CollierCameron10} measured a sky-projected spin-orbit misalignment of $\lambda=-105.8^{\circ}\pm1.2^{\circ}$ 
(using their dataset from McDonald Observatory). They also found an orbital period of $P=1.2198669\pm0.0000012$ days. 
Since WASP-33 b is on a highly inclined, short-period orbit about a rapidly rotating (and therefore likely dynamically oblate) star, \cite{Iorio11} estimated that the orbital nodes should precess at a rate of $d\Omega/dt\leq8.2\times10^{-10}$ s$^{-1}$ ($\leq1.5^{\circ}$ yr$^{-1}$). They predicted that this would result in a changing transit duration that would be detectable in $\sim10$ years. Such a measurement will be challenging, however, as WASP-33 is a $\delta$ Sct variable \citep{Herrero11}. The stellar non-radial pulsations cause distortions in the transit light curve, which could induce systematic errors in the measurement of the transit duration. This change in the transit duration, however, is caused by the changing impact parameter (denoted $b$), which can be more accurately measured using Doppler tomography than using the transit lightcurve. For instance, \cite{CollierCameron10} measured the impact parameter of WASP-33~b to be $b=0.176\pm0.010$ using their spectroscopic data and $b=0.155_{-0.120}^{+0.100}$ using their photometric data. 

It has now been more than six years since the Doppler tomographic observations presented by \cite{CollierCameron10} were obtained. This offers a sufficient time baseline to allow the detection of the movement of the transit chord due to nodal precession. We have thus collected a second epoch of Doppler tomographic observations, and have detected the changing transit chord. 

Orbital precession has previously been detected for Kepler-13 Ab by \cite{Szabo12}, who measured a rate of change of the impact parameter of $db/dt=-0.016\pm0.004$ yr$^{-1}$ using the changing transit duration in {\it Kepler} photometry. Like WASP-33 b, Kepler-13 Ab is a hot Jupiter orbiting a rapidly rotating star on an inclined orbit \citep{Kepler13Ab}. 
\cite{Barnes13} proposed a large rate of nodal precession for the young hot Jupiter candidate PTFO 8-8695 b \citep{vanEyken12}, 
but this planet candidate is still unconfirmed \citep{Ciardi15}. Our measurement of the orbital precession of WASP-33 b is thus the second such measurement for a confirmed exoplanet orbiting a single star. 

\section{Observations and Methodology}

\subsection{Spectroscopic Observations and Analysis}

\cite{CollierCameron10} observed one transit of WASP-33 b with the 2.7m Harlan J. Smith Telescope (HJST) at McDonald Observatory on 2008 November 12 UT, and we reanalyzed these data. They also observed two other transits with other facilities, but we did not reanylize these data. We observed a second transit with the HJST on 2014 October 4 UT, 2,152 days (1,764 planetary orbits) after the first dataset. We obtained both datasets using the Robert G. Tull Coud\'e Spectrograph \citep[TS23;][]{Tull95}, with a spectral resolving power of $R=60,000$. The exposure length was 900 seconds for both datasets. We obtained 13 spectra in 2008 and 21 in 2014; 10 spectra in each dataset were taken during the transit. 

Our methodology for preparing the time series spectra for Doppler tomographic analysis was substantially the same as that described in \cite{Kepler13Ab}. We extracted an average line profile from each spectrum using least squares deconvolution \citep{Donati97}, and subtracted the average out-of-transit line profile to produce time series line profile residuals, which are most useful for analysis. 
 Thanks to WASP-33's brightness ($V=8.3$) we were able to obtain very high quality line profiles; the standard deviation of the continuum was 0.010 of the depth of the line profile for the 2008 observations and 0.0078 for the 2014 dataset. We modeled the line profile perturbation due to the transiting planet by numerically integrating the line profile from each surface element on the star over the visible stellar disk, accounting for the finite exposure time. 

The stellar non-radial pulsations caused a pattern of striations in the time series line profile residuals, complicating the analysis. In order to minimize this effect we exploited the fact that the pulsations propagate in the prograde direction, whereas the planetary orbit is retrograde ($|\lambda|>90^{\circ}$). The frequency components due to the pulsations and the planetary transit thus tended to be separated in the two-dimensional Fourier transform of the time series line profile residuals. 
We constructed a Fourier filter by multiplying each complex element of the Fourier transform 
 by unity if that element was in a region where there was power only from the transit signature, and zero if the element was in a region with significant power from the pulsations, with a Hann function transition between the two regimes. We then performed an inverse Fourier transform on the filtered Fourier spectrum. This successfully removed most of the effects of the pulsations. For best results we had to filter out low-frequency modes where there was power from both the pulsations and the transit; however, the high-frequency components were sufficient to reconstruct most of the transit signature. 

We obtained best-fit values of the transit parameters by exploring the likelihood space of model fits to the data using a Markov chain Monte Carlo (MCMC) with affine-invariant ensemble samplers \citep{GoodmanWeare10}, as implemented in \textsc{emcee} \citep{emcee}. We performed a joint fit to the time series line profile residuals from both 2008 and 2014, as well as a single spectral line, the latter in order to measure the $v\sin i_{\star}$ of WASP-33. For the single line we fit a rotationally broadened line profile to the Ba \textsc{ii} line at 6141.7~\AA, chosen because it is deep but unblended and unsaturated. In order to minimize the impact of the line profile variations we stacked all of our 2008 spectra. We did not utilize the 2014 dataset because we could not obtain a good fit to the Ba \textsc{ii} line.  
 For the time series line profile residuals we fit a transit model computed as described above and passed through our Fourier filter. The MCMC had sixteen parameters: $\lambda$, $b$, and the transit epoch $T_0$ at the two epochs, $v\sin i_{\star}$, $R_p/R_{\star}$, $a/R_{\star}$, $P$, four quadratic limb darkening parameters (two each for the single-line and the time series line profile residual data), the width of the Gaussian line profile from an individual surface element (due to intrinsic broadening, thermal broadening, and microturbulence), and a velocity offset between the single line data and the rest frame. All parameters except $T_0$, $\lambda$ and $b$ were assumed to remain constant between 2008 and 2014. We calculated $T_0$ for 2008 from the epoch and period given by \cite{Kovacs13}, while the 2014 transit epoch was taken from our simultaneous photometric observations of that transit (see \S\ref{phot}). Rather than fitting the limb darkening coefficients directly, we used the triangular sampling method of \cite{Kipping13}. We set Gaussian priors upon $R_p/R_{\star}$, $a/R_{\star}$, $P$, and the 2008 $T_0$, and set the prior value and width to the parameter value and uncertainty, respectively, found by \cite{Kovacs13}, while the prior value and width for the 2014 $T_0$ were taken from our photometric observations. For the limb darkening parameters and prior values we used the same methodology as \cite{Kepler13Ab}.
We used a set of 100 walkers, ran each one for 1000 steps, and cut off the first 500 steps of convergence and burn-in, 
resulting in 50,000 samples from the posterior distribution.

\subsection{Photometric Observations and Analysis}
\label{phot}

Deviation of the observed transit midpoint from that expected based on the published ephemeris could masquerade as a change of the transit chord in the Doppler tomographic data. With exposure lengths of 900 seconds, the spectroscopic data alone did not sufficiently constrain the transit epoch. 
In order to better constrain this parameter we simultaneously obtained photometry of WASP-33 using the Las Cumbres Observatory Global Telescope Network \citep[LCOGT;][]{Brown13} 1m telescope and SBIG camera at McDonald Observatory.

We observed in the Sloan {\it i'} band, and defocused the telescope in order to reduce the effects of inter-pixel variations and avoid saturating on this bright star ($V=8.3$). We obtained 700 images, each with an exposure length of 10 seconds. We used the astrometry.net code \citep{astrometry.net} to register all images, and then performed aperture photometry on WASP-33 and three reference stars using the IDL task APER. We calculated the formal uncertainty on each data point incorporating the uncertainty from APER (based upon photon-counting noise and uncertainty in the sky background), as well as an estimated contribution from the calibration frames and from scintillation noise \citep{Young67}.

We produced a model of the transit lightcurve using the JKTEBOP package\footnote{http://www.astro.keele.ac.uk/jkt/codes/jktebop.html} \citep[e.g.,][]{Southworth11}. We used an MCMC to produce posterior probability distributions for each of the model parameters \citep[we used a custom MCMC derived from that in][not that built into JKTEBOP]{Kepler13Ab}. 

There were significant distortions of the lightcurve due to the aforementioned stellar variability. We treated the stellar variability as correlated noise, and modeled it non-parametrically using Gaussian process regression \citep[e.g.,][]{Gibson12,Roberts13}. This methodology has been used previously to treat stellar noise in transit lightcurves \citep[e.g.,][]{Barclay15}. 

We constructed the covariance matrix $\mathit{K}$ using a Matern 3/2 kernel, where each element of the matrix was
\begin{equation}
k_{ij}=\alpha^2\bigg(1+\frac{\sqrt{3}|t_i-t_j|}{l}\bigg)\exp\bigg(-\frac{\sqrt{3}|t_i-t_j|}{l}\bigg)+\delta_{ij}\sigma_{i}^2
\end{equation}
where $i,j$ denote two of the photometric observations, $t_i$, $t_j$ are the times at which observations $i,j$ were obtained, $\sigma_i$ is the formal uncertainty on datapoint $i$, and $\alpha$ and $l$ are hyperparameters describing the amplitude and timescale of the stellar variability, respectively.

The MCMC used eight parameters: $b$, $R_p/R_{\star}$, $a/R_*$, $\alpha$, $l$, the epoch of transit center $T_{0}$, and two quadratic limb darkening parameters. We set Gaussian priors upon $R_p/R_*$ and $a/R_{\star}$, using the best-fit values and uncertainties found by \cite{Kovacs13} as the center and width of the priors, respectively. We also set priors upon the limb-darkening parameters, using JKTLD\footnote{http://www.astro.keele.ac.uk/jkt/codes/jktld.html} to find the expected limb darkening values from ATLAS model atmospheres from \cite{Claret04} at the stellar parameters of WASP-33 found by \cite{CollierCameron10}. 

\section{Results}

\begin{figure}
\plotone{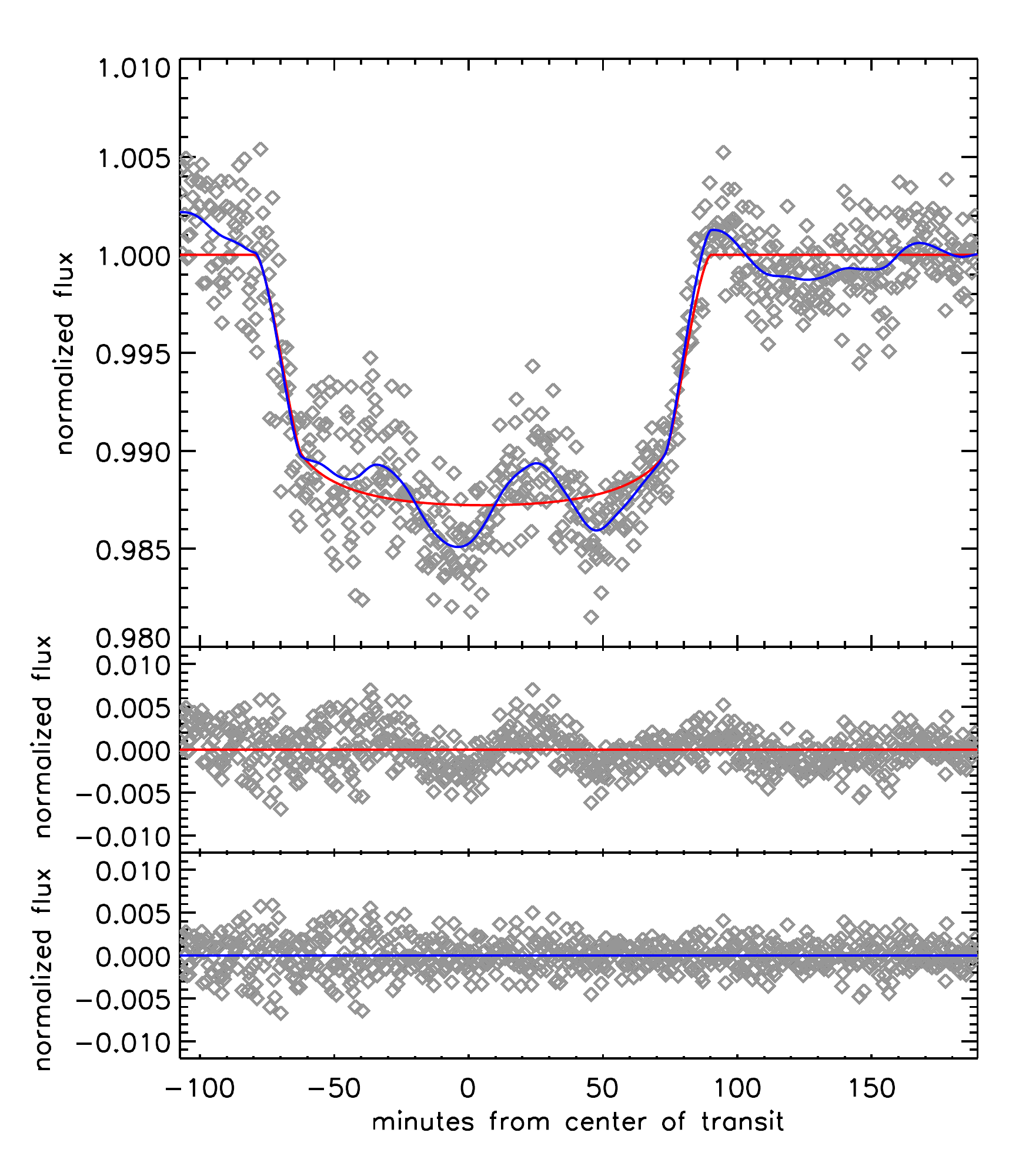}
\caption{Top: LCOGT lightcurve of the transit of WASP-33~b on 2014 October 4 UT. The data are shown in gray, with the best-fit transit model in red and the best-fit transit plus Gaussian process model in blue. Middle: residuals with the best-fit transit model subtracted, showing the stellar variability. Bottom: residuals with the best-fit transit plus Gaussian process model subtracted, showing the power of the Gaussian process to model and remove the stellar variability.  \label{LCfig}} 
\end{figure}

The LCOGT lightcurve is shown in Fig.~\ref{LCfig}.
 We found a best-fit time of transit center of $T_{0}=2456934.77146\pm0.00059$ BJD. This is 12.3 minutes later than predicted by the ephemeris of \cite{CollierCameron10}, but is in agreement with that predicted by the ephemeris of \cite{Kovacs13}. 

\begin{figure*}
\includegraphics[scale=0.5]{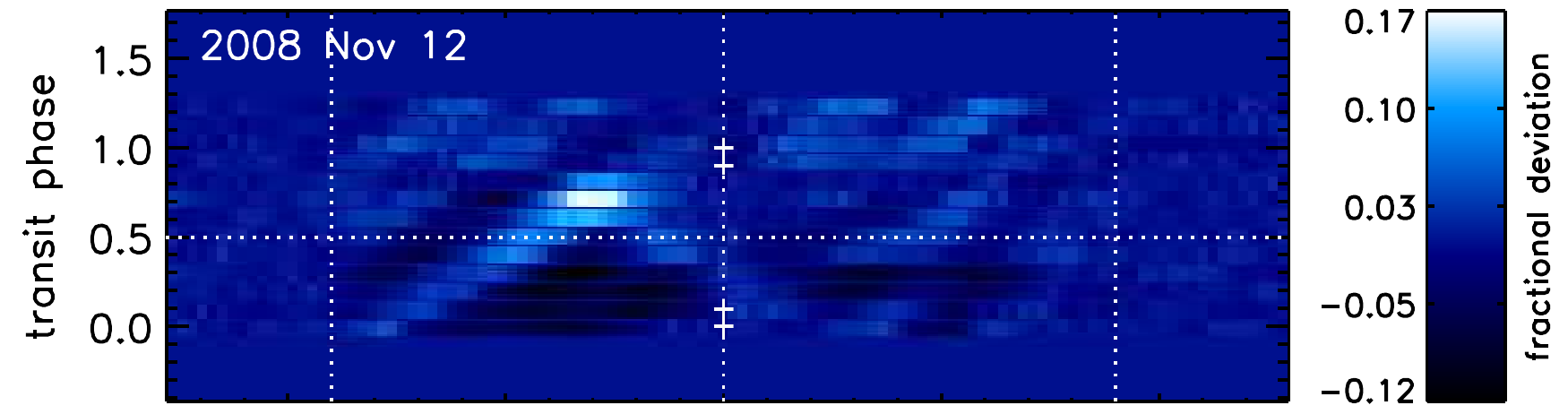}
\includegraphics[scale=0.5]{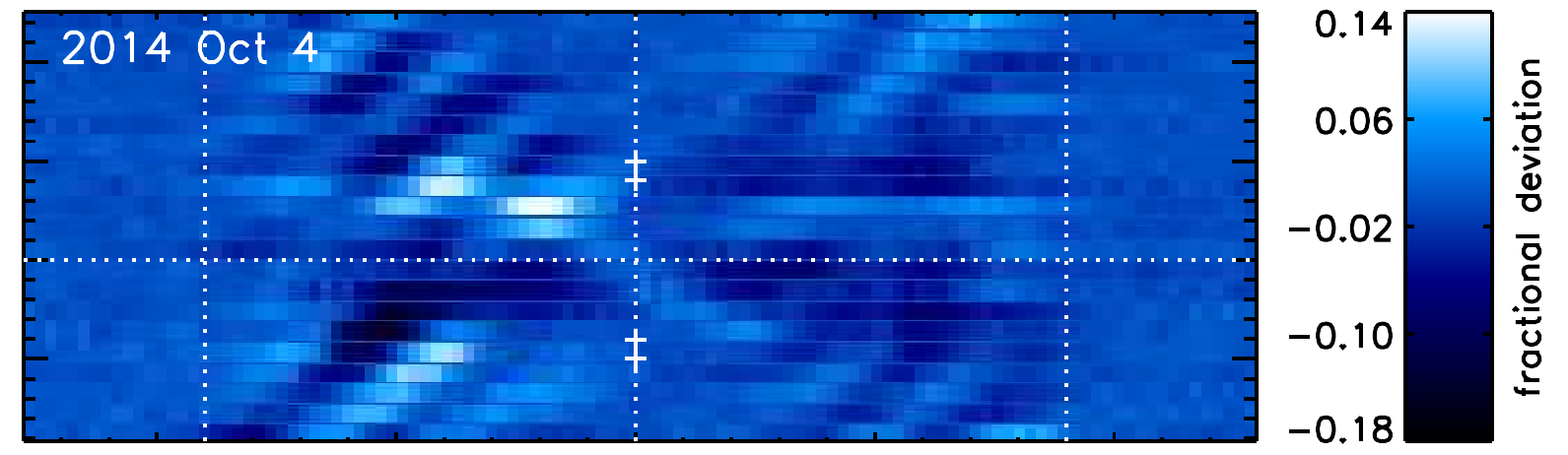}
\includegraphics[scale=0.5]{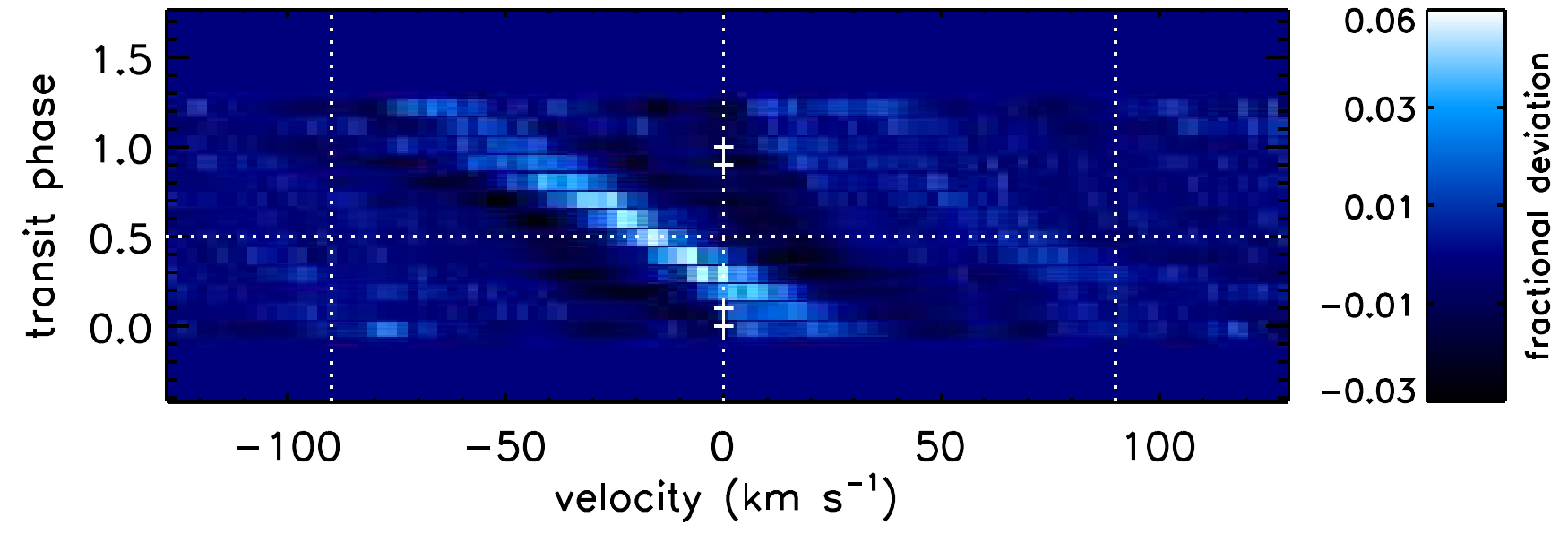}
\includegraphics[scale=0.5]{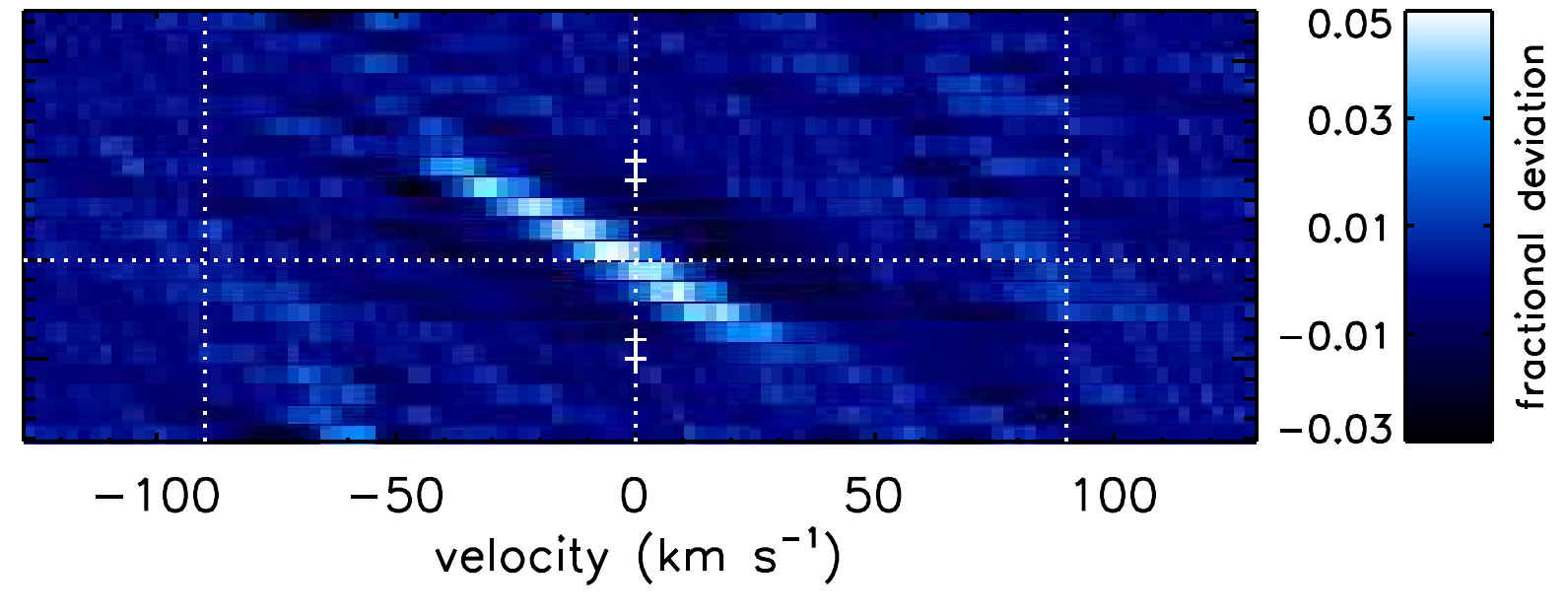}
\includegraphics[scale=0.5]{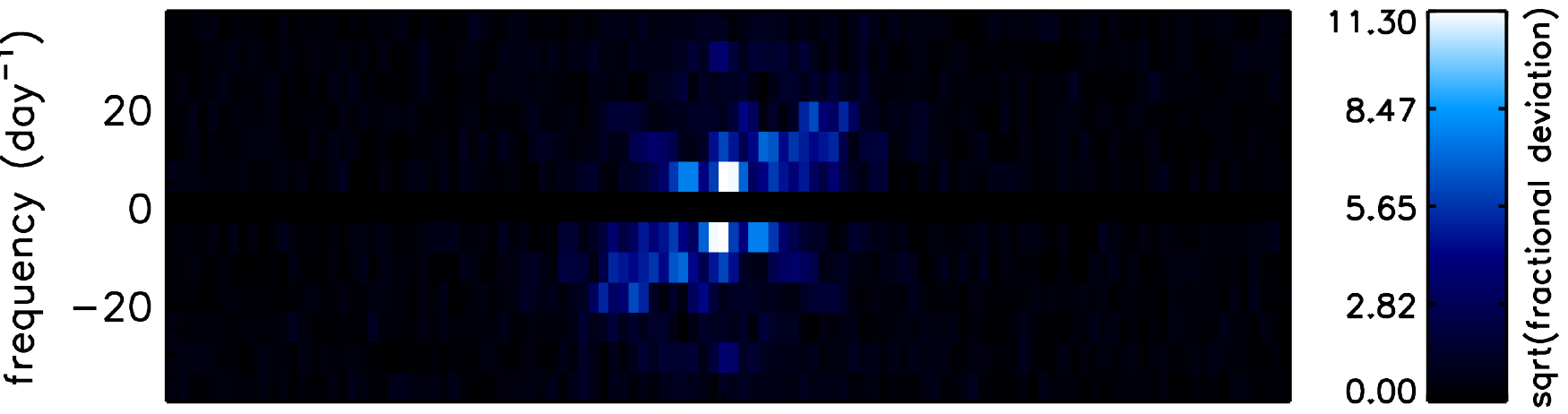}
\includegraphics[scale=0.5]{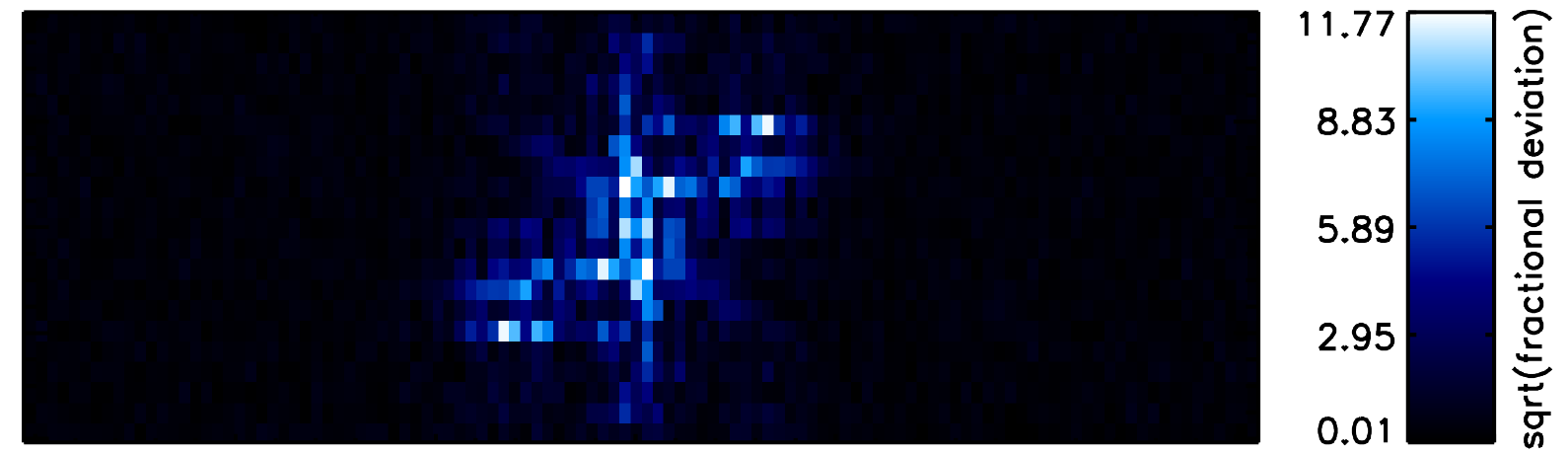}
\includegraphics[scale=0.5]{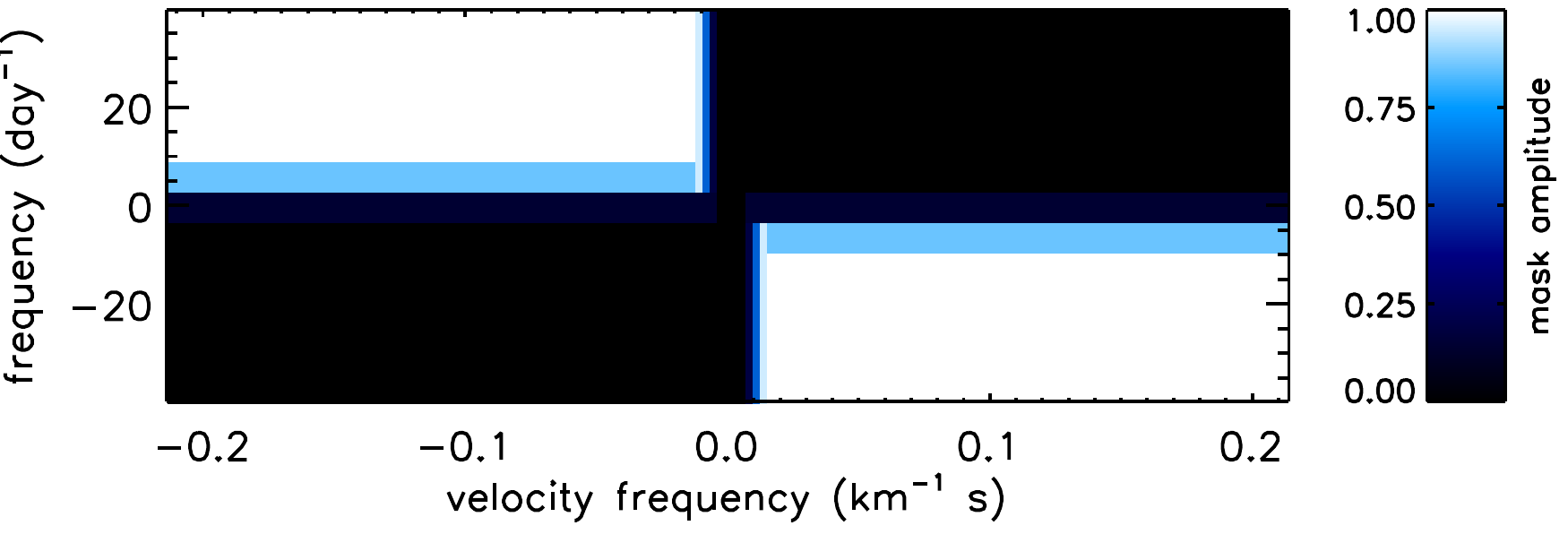}
\hspace{26pt}
\includegraphics[scale=0.5]{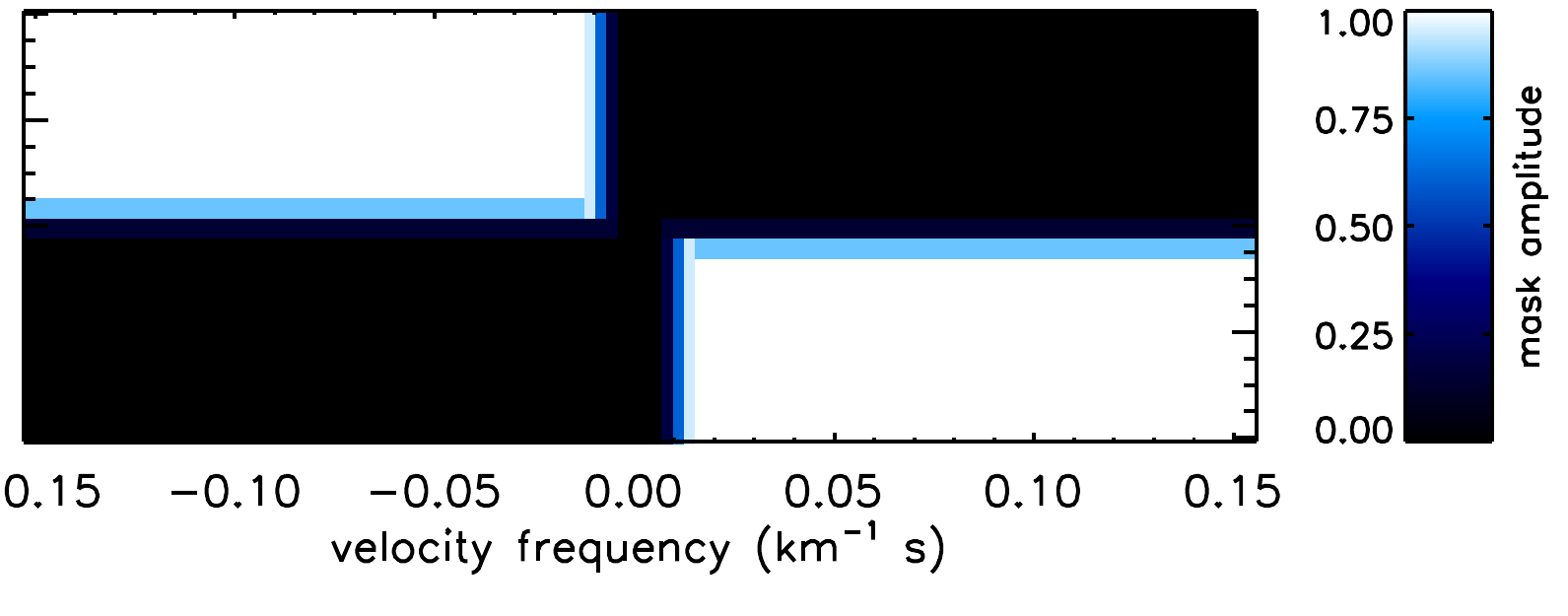}
%\plottwo{wasp33-08maskrev.eps}{wasp33-14mask.eps}
\caption{Doppler tomographic datasets and Fourier filters. The left and right columns show the 2008 and 2014 datasets, respectively. Top row: raw observations. Second row: the data after application of the Fourier filter. Each colorscale line denotes the deviation of the line profile at that time from the average out of transit line profile; bright areas denote shallower areas of the line. Time increases from bottom to top, and we define the ``transit phase'' such that it equals 0 at ingress and 1 at egress. Vertical dashed lines mark $v=0, \pm v\sin i_{\star}$, a horizontal dashed line marks the time of mid-transit, and the four small crosses denote the times of first through fourth contacts. The transit signature is the bright streak running from bottom center towards the upper left. The time range depicted is the same for all plots; flat blue areas indicate regions where we do not have any observations. Most of the remaining anomalous structure in the filtered datasets is ringing due to the filter. The transit signature has shifted slightly to the right between the 2008 and 2014 epochs. Third row: two-dimensional Fourier transforms of the time series line profile residuals, shown with a square-root color scale to best display the frequency structure. The transit signature is the narrow structure running from upper left to lower right. Bottom row: masks used to Fourier filter the data. \label{TSLPRfig}} 
\end{figure*}

We show the time series line profile residuals in Fig.~\ref{TSLPRfig}. 
The best-fit values of the model parameters are given in Table~\ref{restable}. 
We found that both the impact parameter and the spin-orbit misalignment have changed between the two epochs: we measured $b=0.218_{-0.029}^{+0.011}$ and $\lambda=-110.06_{-0.47}^{+0.40}$~$^{\circ}$ in 2008, and $b=0.0840_{-0.0019}^{+0.0020}$ and $\lambda=-112.93_{-0.21}^{+0.23}$~$^{\circ}$ in 2014. 
Our uncertainties on these values are rather small \citep[cf.][whose uncertainties on $\lambda$ and $b$ are $\sim2-5$ times the size of ours]{CollierCameron10}. They did not remove the stellar pulsations from their data, and so it is perhaps not unexpected that we can obtain a more precise result. There may, however, be sources of systematic errors which were not taken into account in our calculation of the uncertainties. 
The values of $\lambda$ and $b$ that we obtained from the 2008 data disagree with those found by \cite{CollierCameron10} by $3.4\sigma$ and $1.3\sigma$, respectively, but agree with another of their Doppler tomographic datasets to within $1.2\sigma$. Most likely this results from differing treatments of the stellar pulsations. Additionally, our posterior distribution for $b$ in 2008 is double-peaked, resulting in asymmetric uncertainties on this parameter. 

Although we did not measure the transit duration directly from our data, we calculated the expected duration using Eqn.~3 of \cite{SeagerMallenOrnelas03}; this is shown in Table~\ref{restable}. The transit duration implied by our Doppler tomographic measurements has changed by 2.7 minutes between the two epochs, a challenging measurement for typical ground-based data, even without the complication of stellar variability.

\begin{figure}
\plotone{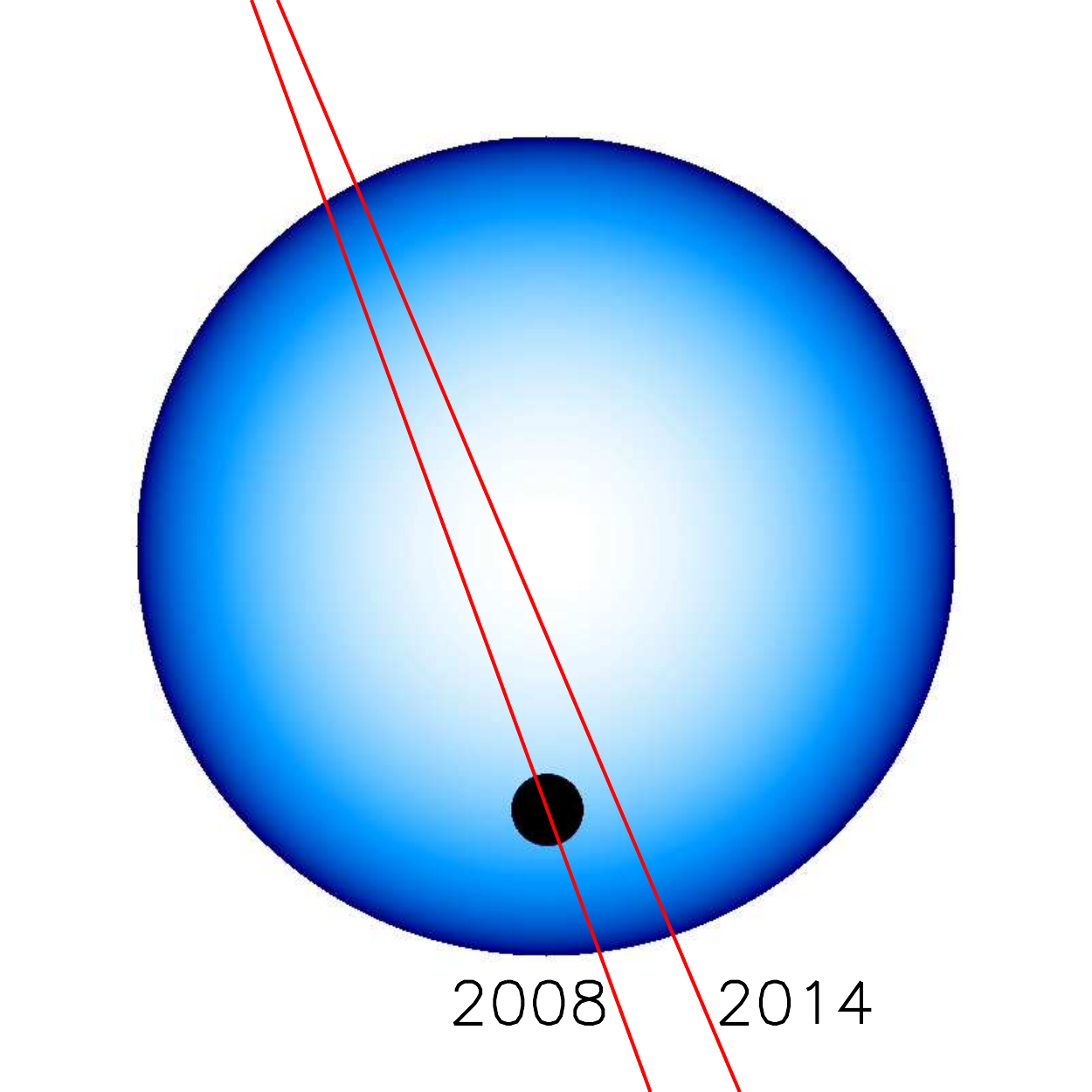}
\caption{Schematic showing the transit chord crossing the star at the 2008 and 2014 epochs. The stellar rotation axis is vertical, and the north pole is at the top, such that star rotates from left to right. The planet moves along the red lines from bottom to top. The silhouette of the planet is shown for a 2008 transit. The limb darkening shown on the star corresponds to that for our best-fit Doppler tomographic solution. Surface brightness variations due to the stellar pulsations are not depicted. \label{schematic}}
\end{figure}

Using our values of $b$ and $\lambda$ at the two epochs, we calculated the rate of precession. We used the definition of the argument of the ascending node $\Omega$ as given by \cite{Queloz00}, i.e., the angle between the plane of the sky and the intersection between the planetary orbital plane and a plane parallel to the line of sight which is also perpendicular to the projection of the stellar rotation axis onto the plane of the sky, as measured in this latter plane. See Fig.~2 of \cite{Queloz00} for a graphical definition; note that the quantity they denoted as $\Delta$ is our $b$, and their $i$ is our $i_{\star}$. 
Using this definition and the definition of the impact parameter, $b=a/R_{\star}\cos i_p$, we related $\Omega$ to our known quantities with
\begin{equation}
\tan\Omega=-\sin\lambda\tan i_p
\end{equation}
which we used to calculate $\Omega$ at the two epochs. We assumed that $a/R_{\star}$ remains constant. We found $db/dt=-0.0228_{-0.0018}^{+0.0050}$ yr$^{-1}$ and $d\lambda/dt=-0.487_{-0.076}^{+0.089}$~$^{\circ}$ yr$^{-1}$, and calculated a rate of nodal precession of WASP-33~b of $d\Omega/dt=0.373_{-0.083}^{+0.031}$~$^{\circ}$ yr$^{-1}$.
This is in agreement with the prediction of \cite{Iorio11}, $d\Omega/dt\leq1.5^{\circ}$ yr$^{-1}$. A schematic view of the changing transit chord is shown in Fig.~\ref{schematic}.

The observation that both $b$ and $\lambda$ are changing implies that the total angular momentum vector of the system $\mathbf{L_{\mathrm{tot}}}$ (the sum of the stellar spin and planetary orbital momentum vectors), about which the planetary orbital angular momentum is precessing, is neither close to perpendicular nor close to parallel to the line of sight. Consider two limiting cases: if $\mathbf{L_{\mathrm{tot}}}$ were perpendicular to the line of sight, then the precession would manifest as purely a change in $b$, whereas if $\mathbf{L_{\mathrm{tot}}}$ were parallel to the line of sight, the precession would manifest as purely a change in $\lambda$. Intermediate motion implies an intermediate angle. 

Using the rate of precession we set limits on the stellar gravitational quadrupole moment $J_2$. This is
\begin{equation}
J_2=-\frac{d\Omega}{dt}\frac{P}{3\pi}\bigg(\frac{a}{R_{\star}}\bigg)^2\sec\psi
\end{equation}
\citep[e.g, from rearranging Eqn. 10 of][]{Barnes13}, where $\psi$ is the angle between the stellar spin and planetary orbital angular momentum vectors ($\lambda$ is $\psi$ projected onto the plane of the sky). This angle can be expressed as \citep[from Eqn. 25 of][]{Iorio11}
\begin{equation}
\cos\psi=\cos i_{\star}\cos i_p+\sin i_{\star}\sin i_p\cos\lambda
\end{equation}
and, while we do not know $i_{\star}$ or $\psi$, following \cite{Iorio11} we can set limits on these quantities. 
 By requiring that WASP-33 rotate at less than the breakup velocity, and using the stellar parameters found by \cite{CollierCameron10}, \cite{Iorio11} set limits of $11.22^{\circ}\leq i_{\star}\leq168.77^{\circ}$. Along with our values of $i_p$ and $\lambda$, this implies a $1\sigma$ range of $93.06^{\circ}\leq\psi\leq110.33^{\circ}$. 
 Thus, we set limits of $9.4\times10^{-5}<J_2<6.1\times10^{-4}$. For comparison, the Solar value is $J_2\sim2\times10^{-7}$ \citep[e.g.,][]{Roxburgh01}.

\begin{deluxetable}{lcc}
\tabletypesize{\scriptsize}
\tablecolumns{3}
\tablewidth{0pt}
\tablecaption{Observed Parameters \label{restable}}
\tablehead{
\colhead{Parameter} & \colhead{2008} & \colhead{2014}
}

\startdata
$v\sin i_{\star}$ (km s$^{-1}$) & $86.63_{-0.32}^{+0.37}$ & \ldots \\
$T_{0}$ (BJD) & \ldots & $2456934.77146 \pm 0.00059$ \\
$\alpha$ & \ldots & $0.00173 \pm 0.00082$ \\
$l$ (minutes) & \ldots & $20.7 \pm 9.2$ \\
$\lambda$ $(^{\circ})$ & $-110.06_{-0.47}^{+0.40}$ & $-112.93_{-0.21}^{+0.23}$ \\
$b$ & $0.218_{-0.029}^{+0.011}$ & $0.0840_{-0.0019}^{+0.0020}$ \\
$i_p$ $(^{\circ})$ & $86.61_{-0.17}^{+0.46}$ & $88.695_{-0.029}^{+0.031}$ \\
$\Omega$ $(^{\circ})$ & $86.39_{-0.18}^{+0.49}$ & $88.584_{-0.032}^{+0.034}$ \\
$\tau_{14}$ (days) & $0.11694_{-0.00041}^{+0.00073}$ & $0.11880\pm0.00033$ \\
\enddata

\tablecomments{Uncertainties are purely statistical and do not take into account systematic sources of error. The transit duration $\tau_{14}$ is calculated from $b$, $P$, and $a/R_{\star}$ and is not measured directly.}

\end{deluxetable}

\section{Conclusions}

We have detected the nodal precession of the hot Jupiter WASP-33 b, and measured a rate of change of the ascending node of $d\Omega/dt=0.373_{-0.083}^{+0.031}$ $^{\circ}$ yr$^{-1}$. This implies that WASP-33 b began transiting its host star as viewed from the Earth in $1974_{-3}^{+8}$, and will transit until $2062_{-10}^{+4}$; the precession period is $\sim970$ years. Furthermore, we have set limits on the stellar gravitational quadrupole moment of $9.4\times10^{-5}<J_2<6.1\times10^{-4}$. 

Given the rate of change of the impact parameter of Kepler-13 Ab found by \cite{Szabo12} and the uncertainty on $b$ measured by \cite{Kepler13Ab}, we expect that the precession of Kepler-13 Ab will be measurable by Doppler tomography by 2017. This will be important as \cite{Masuda15} found that such a measurement will be able to unambiguously distinguish between the conflicting values of the system parameters found by \cite{Kepler13Ab} via Doppler tomography and \cite{Barnes11} using the effects of stellar gravity darkening on the lightcurve. 
No other known planet is currently amenable to the detection of orbital precession with Doppler tomography, as all other planets with published Doppler tomographic observations have approximately aligned orbits and thus should display much slower precession than WASP-33 b \citep{CC189733,Miller10,Gandolfi12,Brown12,Bieryla15,Bourrier15}.
 Current and future transit surveys can provide more targets amenable for the detection of precession via Doppler tomography over the next decade. 

\vspace{12pt}

M.C.J.\ thanks Rapha\"elle Haywood and Daniel Foreman-Mackey for productive discussions introducing him to Gaussian process regression, and the latter for technical suggestions regarding implementation in the MCMC. Thanks to Lorenzo Iorio for his careful reading of and suggestions regarding the draft manuscript, to Ken Rice for initiating a conversation which led to the discovery of an error in the calculation of $J_2$, to Barbara McArthur and Rory Barnes for useful conversations, and to Michael Endl for observing the 2008 transit. M.C.J.\ is supported by a NASA Earth and Space Science Fellowship under Grant NNX12AL59H. This work was also supported by NASA Origins of Solar Systems Program grant NNX11AC34G to W.D.C. This paper includes data taken at The McDonald Observatory of The University of Texas at Austin. This work makes use of observations from the LCOGT network.

\end{document}